\documentclass[conference,a4paper]{APSIPA2025}
\usepackage{amsmath}
\usepackage{graphicx}
\usepackage{multirow}
\usepackage{hyperref}
\usepackage{threeparttable}
\usepackage{amssymb}
\usepackage{amsfonts}
\usepackage{algorithmic}
\usepackage{textcomp}
\usepackage{xcolor}
\usepackage{booktabs}
\usepackage{array}
\usepackage[backend=biber,style=ieee,]{biblatex}
\usepackage{geometry}
\usepackage{fancyhdr}

\fancypagestyle{firststyle}{
  \fancyhf{}
  \fancyhead[C]{2025 Asia Pacific Signal and Information Processing Association Annual Summit and Conference (APSIPA ASC)}
}

\begin{document}

\title{Expressive Prompting: Improving Emotion Intensity and Speaker Consistency in Zero-Shot TTS}

\author{
\IEEEauthorblockN{
Haoyu Wang\authorrefmark{1}, Chunyu Qiang\authorrefmark{1}\authorrefmark{2}, Tianrui Wang\authorrefmark{1}, Cheng Gong\authorrefmark{3}, Yu Jiang\authorrefmark{1},\\ Yuheng Lu\authorrefmark{1}, Chen Zhang\authorrefmark{2}, Longbiao Wang\authorrefmark{1}\authorrefmark{5}, and Jianwu Dang\authorrefmark{4}
}
\authorblockA{
\authorrefmark{1}
Tianjin Key Laboratory of Cognitive Computing and Application, College of Intelligence \\and Computing, Tianjin University, Tianjin, China}

\authorblockA{
\authorrefmark{2}
Kuaishou Technology Co., Ltd, Beijing, China}

\authorblockA{
\authorrefmark{3}
Institute of Artificial Intelligence, China Telecom, China}

\authorblockA{
\authorrefmark{4}
Shenzhen Institute of Advanced Technology, Chinese Academy of Sciences, Guangdong, China}

\authorblockA{
\authorrefmark{5}
Huiyan Technology (Tianjin) Co., Ltd, Tianjin, China}

Corresponding author: longbiao\_wang@tju.edu.cn
}

\maketitle
\thispagestyle{firststyle}
\pagestyle{fancy}
\begin{abstract}
Recent advancements in speech synthesis have enabled large language model (LLM)-based systems to perform zero-shot generation with controllable content, timbre, speaker identity, and emotion through input prompts. As a result, these models heavily rely on prompt design to guide the generation process. However, existing prompt selection methods often fail to ensure that prompts contain sufficiently stable speaker identity cues and appropriate emotional intensity indicators, which are crucial for expressive speech synthesis. To address this challenge, we propose a two-stage prompt selection strategy specifically designed for expressive speech synthesis. In the static stage (before synthesis), we first evaluate prompt candidates using pitch-based prosodic features, perceptual audio quality, and text-emotion coherence scores evaluated by an LLM. We further assess the candidates under a specific TTS model by measuring character error rate, speaker similarity, and emotional similarity between the synthesized and prompt speech. In the dynamic stage (during synthesis), we use a textual similarity model to select the prompt that is most aligned with the current input text. Experimental results demonstrate that our strategy effectively selects prompt to synthesize speech with both high-intensity emotional expression and robust speaker identity, leading to more expressive and stable zero-shot TTS performance. Audio samples and codes will be available at \url{https://whyrrrrun.github.io/ExpPro.github.io/}.
\end{abstract}

\section{Introduction}

In recent years, speech synthesis technology has made remarkable progress, with the quality of synthesized speech continuously improving. The GPT model~\cite{radford2018improving} has achieved great success in the field of natural language processing. Inspired by this, language models have gradually been introduced into the field of speech synthesis and have become the mainstream framework paradigm~\cite{seedtts,wang2023neural,wang2023viola,du2024cosyvoice}. The quality of synthesized speech has progressively reached a level comparable to that of human speech. LM-based TTS systems utilize neural audio codecs~\cite{hsu2021hubert,qiang2024learning,defossez2022high,du2024cosyvoice} to convert speech into discrete tokens, which encapsulate extensive information about the speech. These systems then employ a language model architecture to generate subsequent speech tokens autoregressively. Existing LM-based TTS models implement in-context learning capabilities.

Current LM-based TTS methods~\cite{wang2023viola,du2024cosyvoice} employ autoregressive generation to produce subsequent tokens from input prompts and text. These advanced TTS methods achieve zero-shot voice cloning with just a few seconds of prompt speech. However, the quality of the prompts significantly influences the generated speech output, impacting aspects such as timbre, perceptual quality, and emotional expression~\cite{kojima2022large, JMLR:v25:23-0870}.

Consequently, selecting an appropriate prompt is crucial~\cite{nashid2023retrieval, wang2022learning, shum2023automatic}. There are two mainstream methods for prompt selection: 1) Random: randomly choosing speech from a specific speaker with a certain emotional speech, or 2) Text-based Methods\cite{lou2023universal,reimers-2019-sentence-bert}: selecting prompts based on the similarity between the synthesized text and prompt text. However, these methods are primarily designed for general scenarios and face limitations in emotional speech synthesis. Random selection often fails to provide rich emotional information and expressive capabilities, and focusing solely on the text can yield subpar emotional performances, as there's frequently a weak connection between the text and the desired emotion~\cite{bott2024controlling, seedtts}. Therefore, additional research is required to identify prompts that can enhance emotional expressiveness, speaker similarity, and stability across various LM-based methods in emotional speech synthesis scenarios~\cite{li2024mm, zmm_tts}.
This observation motivates our central research question:
\emph{Can we improve emotion intensity and speaker consistency in zero-shot TTS through better prompt selection, without any additional training?}

\begin{figure*}[t]
  \centering  
  \vspace{-0.3cm}
  
  \includegraphics[width=\linewidth,clip,trim=0 29pt 0 0]{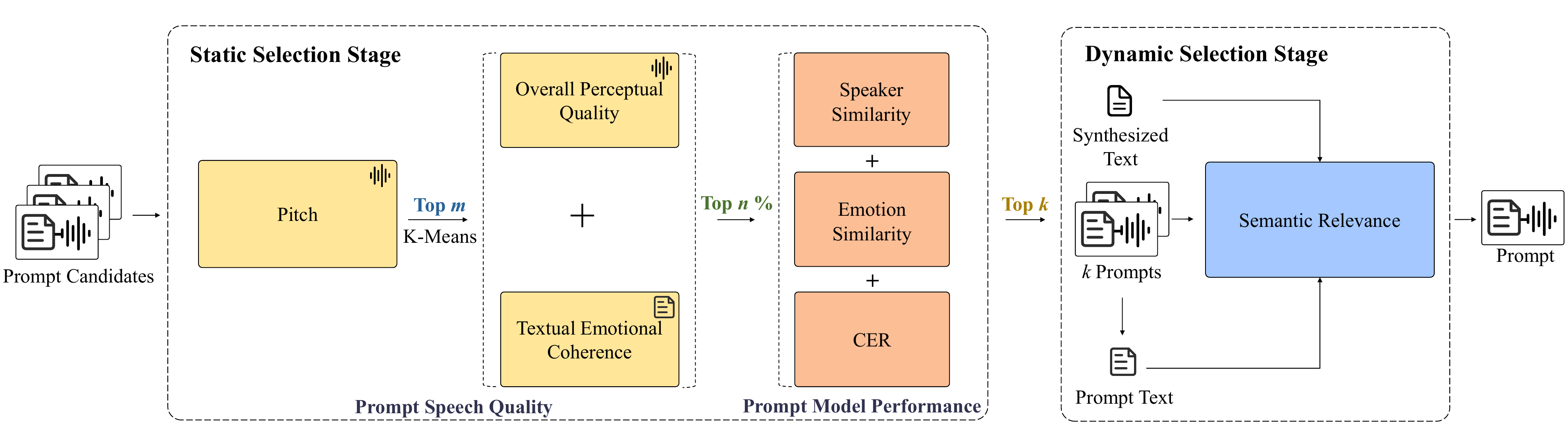}
  \caption{The overview of ExpPro. It consists of two stages: a static selection stage and a dynamic selection stage. The static selection stage evaluates the intrinsic quality of the prompt and its performance in the specific LM-based model, while the dynamic selection stage chooses the most relevant prompt from \( k \) prompts based on the synthesized text.}  
  \label{fig:overview}  
  \vspace{-0.5cm}
\end{figure*}

To tackle these challenges, we propose an innovative two-stage prompt selection strategy --- ExpPro. In the static selection stage, we evaluate both the inherent emotional quality of the prompt candidates and their specific expressive power within the model. In the dynamic selection stage, we choose the most semantically relevant and contextually appropriate prompts from the candidates after the static selection stage, based on the synthesized text. This strategy aims to systematically screen and rank prompts based on various metrics, ultimately selecting prompts with strong emotional expressiveness, high speaker similarity, and high stability. The specific contributions of this paper are as follows:

\begin{enumerate}
    \item We propose a two-stage emotion prompt selection strategy --- ExpPro, which combines static-dynamic selection for LM-based TTS without any additional training.
    \item We conduct a multi-perspective analysis about the text and speech of the prompt, taking into account the ability of prompt in specific models as well as the emotional quality of the prompt itself.
    \item This is a flexible prompt selection strategy suitable for improving emotional expressiveness in any TTS models that involve the concept of prompt speech.
\end{enumerate}

\section{Method}

\subsection{Overview}

The proposed ExpPro is illustrated in Fig.~\ref{fig:overview}, and it consists of two stages: static selection and dynamic selection. In the static selection stage, we select prompt candidates based on emotional expressiveness, perceptual quality, and textual emotional coherence. The selected candidates are then used for inference with the LM-based TTS methods. The objective metrics are used to evaluate candidates and retain those with high quality, expression, and stability. In the dynamic selection stage, we identify the prompt with the highest semantic relevance to the synthesized text input, choosing from the previously filtered candidates. Finally, this prompt is the one that best reflects the required emotional effect of the synthesized text under the current model.

\subsection{Static Selection}

For prompt static selection, we evaluate the quality of the prompt speech across three key dimensions: pitch, overall perceptual quality, and textual emotional coherence derived from a large language model~\cite{achiam2023gpt}. Additionally, we assess the inference results of the prompt candidates, considering metrics such as character error rate (CER), emotion similarity (ES), and speaker similarity. By integrating these factors, we identify the prompt candidates deemed most suitable for the emotion.

\subsubsection{Pitch}

The pitch, or fundamental frequency, is a preverbal feature that imparts tonal and rhythmic qualities to speech~\cite{rodero2011intonation}. As a suprasegmental speech feature, pitch conveys information over a longer time scale than segmental features such as spectral envelopes. Features describing overall attributes of the pitch contour, such as mean and variance, are more emotionally resonant than those describing the pitch shape itself, such as slope, curvature, and inflection~\cite{busso2009analysis}.
a) \textbf{Mean}: This refers to the average pitch level over a period of speech. It can indicate the general tone or mood of the speaker.
b) \textbf{Variance}: This measures the variability in pitch over time. Greater variance might suggest more animated or emotional speech, while less variance could indicate a monotone delivery.

\begin{figure}[t]  
  \centering  
  \includegraphics[width=0.9\linewidth,clip,trim=0 20pt 0 0]{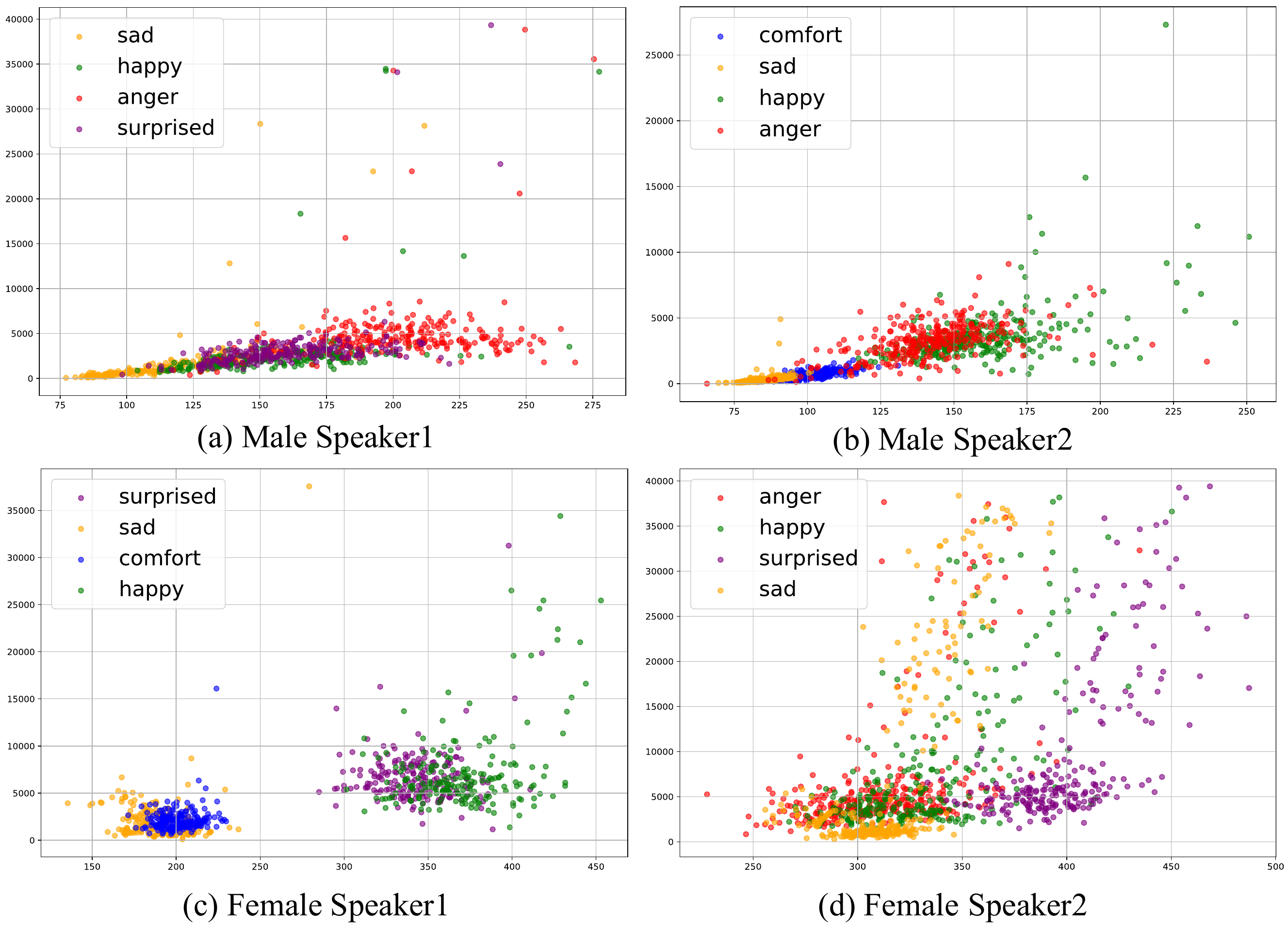}  
  \caption{Mean and variance of emotional speech pitch: red indicates angry, blue indicates comfort, orange indicates sad, green indicates happy, and purple indicates surprised. The x-axis represents the pitch mean, and the y-axis represents the pitch variance.}  
  \label{fig:pitch}  
  \vspace{-0.5cm}
\end{figure}
Different emotional states are associated with distinct pitch patterns~\cite{gharavian2010pitch}. Both sadness and comfort exhibit relatively low mean and variance in pitch, indicating calmer and lower pitch characteristics, with sadness being slightly more subdued. On the other hand, emotions like happiness and surprise demonstrate higher mean and variance, reflecting more pronounced emotional intensity~\cite{frick1985communicating}. Fig. \ref{fig:pitch} illustrates the mean and variance of pitch across various emotional audio samples.

\begin{table*}[h]
\centering
\vspace{-0.3cm}
\caption{The prompt settings for the ChatGPT in the Textual Emotional Coherence module.}
\begin{tabular}{@{}p{2\columnwidth}@{}}
\toprule
\hspace{2em}\textit{You are tasked with evaluating the degree of match between the input text and the emotional label. I will provide the input text and the corresponding emotional label, and you need to assess the degree of match between them. Please note that your response should be a specific score with four decimal places, without any explanation.} \\
\hspace{2em}\textit{For example:} \\
\hspace{4em}\textbf{Text}: \textit{The kitten has been gone for many days, and thinking about it still makes me very sad.} \\
\hspace{4em}\textbf{Emotion}: \textit{Sad.} \\
\hspace{4em}\textbf{Your output}: \textit{0.9357.} \\

\hspace{2em}\textit{Now, here's my formal question:}\\
\hspace{4em}\textbf{Text}: {[}Text{]}, \textbf{Emotion}: {[}Emotion{]}. \textbf{Your output}:\\
\bottomrule
\end{tabular}
\label{tab:prompt}
\vspace{-0.5cm}
\end{table*}

We select the prompt speech based on the distinct tonal features associated with each emotion category. Initially, we calculated the mean and variance for each emotion type. Subsequently, we apply the K-Means algorithm to cluster 10 groups based on the mean and variance of the prompt candidates for different speakers and emotions. We select \( m \) clusters with stronger or weaker means and variances based on the various states of different emotional classes.

\subsubsection{Perceptual and Textual Selecting}
We comprehensively consider both perceptual quality and text consistency.


\textbf{Overall Perceptual Quality}: We regard the quality of the prompt speech as a critical factor and utilize DNSMOS~\cite{reddy2021dnsmos} for this purpose. DNSMOS is a deep learning-based audio quality assessment tool designed to evaluate the quality of audio signals. It can assess the clarity, naturalness, and overall quality of audio. Leveraging neural network models to simulate human auditory perception, it provides objective scores that are highly correlated with subjective ratings. We measure DNSMOS on all results following pitch selection.

\textbf{Textual Emotional Coherence}:
When the text of speech aligns more closely with a particular emotional expression, the sentence can more effectively convey the desired emotion. To assess the relevance of the text to the corresponding emotion in the prompt speech, we use the ChatGPT\footnote{https://chatgpt.com/} API. While ChatGPT operates as a black-box system, we use it not as a definitive classifier, but as a semantic comparator to assess the relative coherence between input text and target emotion in a controlled and consistent setting.

Specifically, we fix a benchmarking prompt (shown in Table~\ref{tab:prompt}) and use ChatGPT under identical conditions for all comparisons. This ensures that all samples are evaluated under the same latent criteria, enabling fair and reproducible ranking across prompt candidates.

Moreover, we verified the stability and reliability of the ChatGPT-based assessment via manual inspection on a sample of cases, which showed strong alignment with human intuition. This setup allows us to benefit from the model's strong contextual understanding while avoiding over-dependence on its opaque internal mechanisms.



Considering the relatively stable distribution of DNSMOS scores within the same dataset, to highlight differences, we directly combine the textual emotional coherence scores with the DNSMOS scores and select the top \( n\% \) as the most emotionally expressive data.

\subsubsection{Selecting with Performance under LM-based TTS Method}

The method above focuses on the selection method for evaluating the quality of the prompt speech itself. Additionally, we recognize that even when identical prompt speech is input into different methods, the resulting outputs can vary significantly. This variability primarily depends on factors such as the selection of speech tokens. To address this question, we propose a strategy that considers the specific performance of different models when processing the same prompt speech.

Specifically, we select 20 neutral descriptive sentences for inference based on the prompt candidates from our prompt speech quality selection process. We then evaluate the inference results for all prompt speeches by calculating the CER of the synthesized speech. Furthermore, we use Resemblyzer~\cite{wan2018generalized} and WavLM~\cite{chen2022wavlm} to evaluate speaker similarity and assess the model's capability to generate the same speaker's voice from the given prompt. Finally, we employ the emotion2vec~\cite{ma2023emotion2vec} model to assess the cosine similarity between the emotion of synthesized speech and prompt speech, which serves as an indicator of the model's effect in capturing the emotional information of the prompt speech.\label{SCM}

The three metrics of CER, speaker similarity, and ES form the framework for assessing our model's effect in capturing various speech information. These metrics are further integrated through a weighted sum to provide a comprehensive assessment. In our selection strategy, we prioritize the quality of the prompt candidates themselves. We start with an initial selection of their intrinsic emotional quality before applying the model-specific selection method.

\subsection{Dynamic Selection}

The consistency between the prompt text and the synthesized text also affects the results, so we employ a dynamic selection strategy based on the text. The stsb-distilroberta-base\footnote{https://huggingface.co/cross-encoder/stsb-distilroberta-base} analyzes the currently synthesized text alongside the statically selected speeches from the prompt candidates. This allows us to identify the most relevant prompt for the current text, which is then chosen as the final prompt.

\section{Experiments}

\vspace{-0.3cm}
\begin{table*}[htbp]
    \centering
    \vspace{-0.3cm}
    \setlength\tabcolsep{4pt}
    \setlength{\extrarowheight}{1pt}
    \caption{Comparison of zero-shot LM-based TTS performance across various prompt selection methods. SP and MOS are presented with a 95\% confidence interval. ESD does not include the comfort emotion, the corresponding position is empty.}
    \begin{tabular}{l|c|c|cc|cc|cc|cc|cc}
    \hline
    \multirow{2}{*}{Model} &
    \multirow{2}{*}{Dataset} &
    \multirow{2}{*}{Method} &
    \multicolumn{2}{c|}{Happy} &
    \multicolumn{2}{c|}{Sad} &
    \multicolumn{2}{c|}{Angry} &
    \multicolumn{2}{c|}{Surprised} &
    \multicolumn{2}{c}{Comfort} \\ \cline{4-13} 
    &     &     & MOS $\uparrow$ & SP $\uparrow$ & MOS $\uparrow$ & SP $\uparrow$ & MOS $\uparrow$ & SP $\uparrow$ & MOS $\uparrow$ & SP $\uparrow$ & MOS $\uparrow$ & SP $\uparrow$ \\ \hline

    \multirow{6}{*}{CosyVoice} &     & Random   &  3.83\textsubscript{±0.12}   &  0.683\textsubscript{±0.023}  &  3.71\textsubscript{±0.09}   &  0.635\textsubscript{±0.019}  &  3.84\textsubscript{±0.13}   &  0.621\textsubscript{±0.011}  &  3.83\textsubscript{±0.19}   &  0.667\textsubscript{±0.015} & - & - \\
                               & ESD & MiniLM   &  3.79\textsubscript{±0.11}   &  0.681\textsubscript{±0.018}  &   3.75\textsubscript{±0.10}  &  0.647\textsubscript{±0.015}  &  3.81\textsubscript{±0.17}   &  0.628\textsubscript{±0.021}  &  3.88\textsubscript{±0.18}   &  0.675\textsubscript{±0.013} & - & - \\
                               &     & \textbf{ExpPro} &  \textbf{3.87\textsubscript{\textbf{±0.14}}}   &  \textbf{0.689\textsubscript{\textbf{±0.012}}}  &   \textbf{3.80\textsubscript{\textbf{±0.13}}}  &  \textbf{0.651\textsubscript{\textbf{±0.014}}}  &  \textbf{3.86\textsubscript{\textbf{±0.12}}}   &  \textbf{0.633\textsubscript{\textbf{±0.015}}}  &  \textbf{3.93\textsubscript{\textbf{±0.16}}}   &  \textbf{0.681\textsubscript{\textbf{±0.014}}} & - & - \\ \cline{2-13}

    &           & Random   &  4.13\textsubscript{±0.11}   &  0.767\textsubscript{±0.018}  &   4.23\textsubscript{±0.16}  &  0.789\textsubscript{±0.023}  &  4.43\textsubscript{±0.16}   &  0.733\textsubscript{±0.018}  &  4.11\textsubscript{±0.11}   &  0.733\textsubscript{±0.013}  &   4.21\textsubscript{±0.11}  &  0.833\textsubscript{±0.013}  \\
    & HE2D & MiniLM   &  4.33\textsubscript{±0.14}   &  0.767\textsubscript{±0.013}  &   4.35\textsubscript{±0.15}  &  0.818\textsubscript{±0.017}  &  4.46\textsubscript{±0.12}   &  0.767\textsubscript{±0.018}  &  4.30\textsubscript{±0.09}   &  0.744\textsubscript{±0.015}  &   4.34\textsubscript{±0.12}  &  0.879\textsubscript{±0.011}  \\
    &           & \textbf{ExpPro} &  \textbf{4.45\textsubscript{\textbf{±0.13}}}   &  \textbf{0.811\textsubscript{\textbf{±0.017}}}  &   \textbf{4.42\textsubscript{\textbf{±0.11}}}  &  \textbf{0.832\textsubscript{\textbf{±0.018}}}  &  \textbf{4.50\textsubscript{\textbf{±0.12}}}   &  \textbf{0.867\textsubscript{\textbf{±0.015}}}  &  \textbf{4.33\textsubscript{\textbf{±0.08}}}   &  \textbf{0.767\textsubscript{\textbf{±0.017}}}  &   \textbf{4.36\textsubscript{\textbf{±0.11}}}  &  \textbf{0.889\textsubscript{\textbf{±0.011}}}  \\ \hline

    \multirow{6}{*}{GPT-SoVITS} &      & Random   &  3.69\textsubscript{±0.18}   &  0.641\textsubscript{±0.021}  &  3.57\textsubscript{±0.13}   &  0.597\textsubscript{±0.015}  &  3.59\textsubscript{±0.12}   &  0.583\textsubscript{±0.017}  &  3.72\textsubscript{±0.11}   &  0.641\textsubscript{±0.013} & - & - \\
                                & ESD  & MiniLM   &  3.71\textsubscript{±0.15}   &  0.637\textsubscript{±0.022}  &   3.62\textsubscript{±0.14}  &  0.628\textsubscript{±0.014}  &  3.51\textsubscript{±0.12}   &  0.591\textsubscript{±0.019}  &  3.77\textsubscript{±0.13}   &  0.671\textsubscript{±0.019} & - & - \\
                                &      & \textbf{ExpPro} &  \textbf{3.78\textsubscript{\textbf{±0.17}}}   &  \textbf{0.659\textsubscript{\textbf{±0.017}}}  &   \textbf{3.75\textsubscript{\textbf{±0.11}}}  &  \textbf{0.641\textsubscript{\textbf{±0.016}}}  &  \textbf{3.65\textsubscript{\textbf{±0.11}}}   &  \textbf{0.611\textsubscript{\textbf{±0.013}}}  &  \textbf{3.88\textsubscript{\textbf{±0.14}}}   &  \textbf{0.675\textsubscript{\textbf{±0.016}}} & - & - \\ \cline{2-13}

                                &      & Random   &  4.02\textsubscript{±0.15}   &  0.668\textsubscript{±0.023}  &   3.98\textsubscript{±0.19}  &  0.727\textsubscript{±0.024}  &  4.01\textsubscript{±0.09}   &  0.696\textsubscript{±0.013}  &  4.12\textsubscript{±0.11}   &  0.711\textsubscript{±0.021}  &   4.09\textsubscript{±0.13}  &  0.789\textsubscript{±0.016}  \\
                                & HE2D & MiniLM   &  4.29\textsubscript{±0.12}   &  0.709\textsubscript{±0.020}  &   4.25\textsubscript{±0.17}  &  0.794\textsubscript{±0.018}  &  4.27\textsubscript{±0.15}   &  0.733\textsubscript{±0.012}  &  4.30\textsubscript{±0.16}   &  0.756\textsubscript{±0.024}  &  4.42\textsubscript{±0.11}   &  0.767\textsubscript{±0.014} \\
                                &     & \textbf{ExpPro} &  \textbf{4.39\textsubscript{\textbf{±0.17}}}   &  \textbf{0.733\textsubscript{\textbf{±0.019}}}  &   \textbf{4.37\textsubscript{\textbf{±0.14}}}  &  \textbf{0.826\textsubscript{\textbf{±0.015}}}  &  \textbf{4.44\textsubscript{\textbf{±0.12}}}   &  \textbf{0.790\textsubscript{\textbf{±0.012}}}  &  \textbf{4.31\textsubscript{\textbf{±0.13}}}   &  \textbf{0.778\textsubscript{\textbf{±0.016}}}  &   \textbf{4.46\textsubscript{\textbf{±0.13}}} &  \textbf{0.811\textsubscript{\textbf{±0.013}}}  \\ \hline

    \end{tabular}
    \label{tab:main}
    \vspace{-0.5cm}
\end{table*}

\subsection{Data}

We utilize a highly emotionally expressive dataset (HE2D), as described in~\cite{qiang2023improving}, which includes recordings from two male and two female speakers to validate our prompt selection strategy. The dataset encompasses five distinct emotions: comfort, happy, sad, angry, and surprise. Each speaker exhibits four of these emotions, with 200 samples per emotion, amounting to a total of 800 samples per speaker. Additionally, to demonstrate the generalizability of our approach, we also validate our method using the open-source Emotional Speech Database (ESD)~\cite{zhou2021seen}. From this dataset, we randomly select recordings from four Chinese speakers (two male and two female), each corresponding to four emotional categories.

\subsection{Compared Methods}

To verify the effectiveness of our approach, we compare the following strategies for selecting prompt speech:
1) \textbf{Random:} We randomly select from all prompts as the prompt choice. 2) \textbf{Text-based Methods:} We achieve the selection by performing semantic similarity analysis between the synthetic text and the prompt text~\cite{reimers-2019-sentence-bert}, using all-MiniLM-L6-v2\footnote{https://huggingface.co/sentence-transformers/all-MiniLM-L6-v2} (MiniLM)~\cite{wang2020minilm} to implement the prompt selection.

\subsection{Test Metrics}

For our subjective evaluation, we employ 20 native speakers. For the main method comparison, we provide 5 sentences per emotion, each designed to convey the corresponding emotional state. For other ablation experiments, we select 5 descriptive neutral sentences to verify the effectiveness of our method. We provide participants with detailed evaluation criteria and report both the mean scores and 95\% confidence intervals. The test metrics used in the subjective evaluation are as follows:
\begin{itemize}
    \item \textbf{Emotion MOS (MOS): }
    This metric evaluates the quality and emotional expression of the synthesized speech.
    \item \textbf{Strength Perception (SP): }
    A subjective strength perception test. The judge is asked to rate the emotional strength on a scale from 0 to 1.
\end{itemize}

The object evaluation metrics include speaker similarity, ES, CER. Resemb and WavLM are calculated via cosine similarity between speaker representations of the target and generated speech using Resemblyzer~\cite{wan2018generalized} and WavLM-large~\cite{chen2022wavlm}, while ES uses cosine similarity between emotion2vec-large~\cite{ma2023emotion2vec} representations. CER compares the synthesized text with Paraformer-zh~\cite{gao2022paraformer} output.

\section{Experimental Results}

In our experiments, we validate our method through multiple approaches. Specifically, we conduct evaluations using two state-of-the-art and widely popular LM-based TTS models: CosyVoice~\cite{du2024cosyvoice} and GPT-SoVITS\footnote{https://github.com/RVC-Boss/GPT-SoVITS}. First, we compare our method with the baseline approach on prompt speech selection during zero-shot inference for both TTS models. Next, we explore the effects of model parameter selection and systematically validate the contribution of each module.

\subsection{Comparison with Baseline Methods}


By applying the prompt selection process during zero-shot inference with these two models, we compare our approach against other baseline methods. Detailed experimental results are presented in Table \ref{tab:main}. Our proposed method demonstrates significantly better performance across major emotional categories compared to the baselines, with noticeable improvements in both emotional intensity and stability. When comparing the results on the ESD dataset with those on the H2ED dataset, we observe that the stronger the emotional expressiveness of the prompt speech, the more pronounced the improvements achieved by our method. This approach thoroughly considers the quality of the prompts, the performance differences between models, and the correlation between synthesized text and prompt text, resulting in superior experimental outcomes.

\begin{table}[htbp]
    \centering
    \vspace{-0.2cm}
    \setlength{\extrarowheight}{1pt}
    \setlength{\tabcolsep}{4pt}
    \caption{The results of ExpPro in different LM-based TTS with the same prompt candidates.}
    \begin{tabular}{l|ccccc}
        \cline{1-6}
        Model & PromptID & CER $\downarrow$& Resemb $\uparrow$& WavLM $\uparrow$& ES$\uparrow$\\
        \cline{1-6}
         & 165(Top1) & 1.55\% & 93.66 & 82.10 & 98.37 \\

         & 112(Top2) & 2.01\% & 90.67 & 81.74 & 98.45 \\

         CosyVoice & 119(Top3) & 1.86\% & 91.68 & 79.17 & 97.61 \\

         & 047(Bottom2) & 2.94\% & 84.43 & 63.67 & 50.30 \\
         
         & 029(Bottom1) & 2.63\% & 89.65 & 69.63 & 43.36 \\
         
        \cline{1-6}

         & 083(Top1) & 1.55\% & 85.82 & 69.11 & 93.11 \\

         & 031(Top2) & 2.01\% & 85.27 & 61.75 & 96.17 \\

         GPT-SoVITS & 064(Top3) & 1.70\% & 87.71 & 61.53 & 93.01 \\
         
         & 099(Bottom2) & 1.75\% & 79.71 & 49.39 & 65.37 \\
         
         & 039(Bottom1) & 2.24\% & 77.40 & 41.89 & 52.44 \\

        \cline{1-6}

    \end{tabular}
    \label{tab:different}
    \vspace{-0.3cm}
\end{table}

\subsection{Evaluation on Different TTS Models}

We further demonstrate the importance of the prompt model performance module. The results, presented in Table \ref{tab:different}, are obtained by ranking the prompts according to the weighted-sum evaluation of CER, speaker similarity, and ES, using the same happy emotion data from female speaker 1 across two different models. Our findings reveal that the Top three most expressive prompts and the Bottom two least expressive prompts differ significantly between the two models, despite utilizing identical input data. This indicates that the effectiveness of a given prompt is model-dependent, underscoring the necessity of evaluating prompt performance within the context of the specific TTS system.

Furthermore, the results indicate that CosyVoice outperforms GPT-SoVITS in terms of both speaker similarity and emotion similarity. This performance advantage is primarily attributed to CosyVoice's use of an ASR-supervised tokenizer and the integration of additional speaker x-vector inputs. Based on this result, all subsequent experiments are conducted primarily using the CosyVoice model.

\begin{table}[htbp]
    \centering
    \setlength{\extrarowheight}{1pt}
    \vspace{-0.15cm}
    \caption{The result of different parameter settings. SP is presented with a 95\% confidence interval.}
    \begin{tabular}{l|cccc}
        \cline{1-5}
        & \( m \) & \( n \) & ES $\uparrow$& SP $\uparrow$\\
        \cline{1-5}
        1 & 5 & 25 & 88.85 & 0.724\textsubscript{±0.013} \\
        2 & 4 & 25 & 89.12 & 0.733\textsubscript{±0.015} \\
        3 & 3 & 25 & \textbf{89.83} & 0.735\textsubscript{±0.015} \\
        4 & 2 & 25 & 89.25 & \textbf{0.737\textsubscript{\textbf{±0.013}}} \\
        \cline{1-5}
        5 & 3 & 25 & 89.83 & 0.735\textsubscript{±0.014} \\
        6 & 3 & 20 & 92.27 & 0.745\textsubscript{±0.016} \\
        7 & 3 & 15 & 94.01 & 0.750\textsubscript{±0.013} \\
        8 & 3 & 10 & \textbf{95.59} & \textbf{0.767\textsubscript{\textbf{±0.011}}} \\
        \cline{1-5}
    \end{tabular}
    \label{tab:KMN}
    \vspace{-0.35cm}
\end{table}

\subsection{Range of Prompt Selection}
We conduct experiments on the range of data selected at each stage, including the selection of \( m, n, k \), and other variables under different conditions. The specific results are shown in Table \ref{tab:KMN}. From the table, we can clearly observe that as the degree of selection of the pitch clusters increases (\( m \) decreases), the emotional impact of the prompts also gradually enhances. The results indicating that emotion similarity and strength perception increase as \( n \) decreases highlight the effectiveness of using overall perceptual quality and textual emotional coherence to prompt selection. Considering the limited number of prompt speeches, the prominence of the prompter's emotional effect, and the uncertainty of text content during inference, we ultimately choose the parameters \( m=3 \), \( n=15 \), and \( k=5 \).

\vspace{-0.2cm}
\begin{table}[htbp]
    \centering
    \setlength{\extrarowheight}{1pt}
    \setlength\tabcolsep{3pt}
    \caption{The experiment of quality selection. $\circleddash$ represents the reverse method of ExpPro, PSQ denotes prompt speech quality. SP is presented with a 95\% confidence interval.}
    \begin{tabular}{l|rccccc}
        \cline{1-7}
        Emotion & Method & CER $\downarrow$& Resemb $\uparrow$& WavLM $\uparrow$& ES $\uparrow$ & SP $\uparrow$\\
        \cline{1-7}
        \multirow{2}{*}{Happy} & ExpPro & \textbf{2.35\%} & \textbf{91.37} & 73.51 & \textbf{93.08} & \textbf{0.782\textsubscript{\textbf{±0.018}}} \\
         & $\circleddash$PSQ & 2.35\% & 91.31 & \textbf{74.75} & 90.46 & 0.633\textsubscript{±0.014} \\
        \cline{1-7}
        \multirow{2}{*}{Sad} & ExpPro & \textbf{2.23\%} & \textbf{90.28} & \textbf{81.41} & 96.31 & \textbf{0.724\textsubscript{\textbf{±0.013}}} \\
         & $\circleddash$PSQ & 2.32\% & 89.90 & 78.03 & \textbf{97.47} & 0.674\textsubscript{±0.017} \\
        \cline{1-7}
        \multirow{2}{*}{Angry} & ExpPro & \textbf{1.83\%} & \textbf{92.33} & \textbf{76.59} & \textbf{96.31} & \textbf{0.697\textsubscript{\textbf{±0.013}}} \\
         & $\circleddash$PSQ & 2.23\% & 86.76 & 67.26 & 91.29 & 0.579\textsubscript{±0.016} \\
        \cline{1-7}
        \multirow{2}{*}{Surprised} & ExpPro & \textbf{1.67\%} & 87.41 & \textbf{77.55} & \textbf{93.61} & \textbf{0.744\textsubscript{\textbf{±0.011}}} \\
         & $\circleddash$PSQ & 1.95\% & \textbf{89.25} & 77.03 & 92.15 & 0.646\textsubscript{±0.017} \\
        \cline{1-7}
        \multirow{2}{*}{Comfort} & ExpPro & \textbf{1.70\%} & 91.69 & \textbf{84.32} & 97.56 & \textbf{0.741\textsubscript{\textbf{±0.008}}} \\
         & $\circleddash$PSQ & 1.70\% & \textbf{92.11} & 82.21 & \textbf{97.59} & 0.688\textsubscript{±0.007} \\
        \cline{1-7}
    \end{tabular}
    \label{tab:psq}
    \vspace{-0.2cm}
\end{table}
\subsection{Ablation Study}
To better understand the contribution of each component in ExpPro, we conduct ablation studies by isolating and evaluating the impact of prompt speech quality and prompt model performance in two stages. The two stages are analyzed separately as follows:

\subsubsection{Importance of Prompt Speech Quality}
Table \ref{tab:psq} presents the results of our experiments on the prompt speech quality module. We employ an inverse selection strategy compared to ExpPro to finish these experiments. Specifically, we select the \( m \) clusters that performed the worst after pitch clustering, along with the Bottom \( n\%\) of data based on overall perceptual quality and textual emotional coherence weighted results. Finally, we compare the performance of the Top \( k \) prompts candidates through prompt model performance, respectively. The results indicate that ExpPro successfully selects emotionally expressive prompts. 

\begin{table}[htbp]
    \centering
    \setlength{\extrarowheight}{1pt}
    \vspace{-0.1cm}
    
    \caption{The experiment of model performance selection. $\circleddash$ represents the reverse method of ExpPro, PMP denotes prompt model performance. SP and MOS are presented with a 95\% confidence interval.}
    \begin{tabular}{l|rcc}
        \cline{1-4}
        Emotion & Method & MOS $\uparrow$& SP $\uparrow$\\
        \cline{1-4}
        \multirow{2}{*}{Happy} & ExpPro & \textbf{4.30\textsubscript{\textbf{±0.13}}} & \textbf{0.787\textsubscript{\textbf{±0.017}}}\\
         & $\circleddash$PMP & 4.27\textsubscript{±0.08} & 0.773\textsubscript{±0.017}\\
        \cline{1-4}
        \multirow{2}{*}{Sad} & ExpPro & \textbf{4.24\textsubscript{\textbf{±0.13}}} & \textbf{0.817\textsubscript{\textbf{±0.012}}}\\
         & $\circleddash$PMP & 4.21\textsubscript{±0.13} & 0.773\textsubscript{±0.017}\\
        \cline{1-4}
        \multirow{2}{*}{Angry} & ExpPro & \textbf{4.33\textsubscript{\textbf{±0.15}}} & \textbf{0.700\textsubscript{\textbf{±0.015}}}\\
         & $\circleddash$PMP & 4.28\textsubscript{±0.14} & 0.677\textsubscript{±0.016}\\
        \cline{1-4}
        \multirow{2}{*}{Surprised} & ExpPro & \textbf{4.12\textsubscript{\textbf{±0.11}}} & \textbf{0.727\textsubscript{\textbf{±0.016}}}\\
         & $\circleddash$PMP & 4.11\textsubscript{±0.13} & 0.723\textsubscript{±0.015}\\ 
        \cline{1-4}
        \multirow{2}{*}{Comfort} & ExpPro & \textbf{4.27\textsubscript{\textbf{±0.15}}} & \textbf{0.800\textsubscript{\textbf{±0.011}}}\\
         & $\circleddash$PMP & 4.22\textsubscript{±0.15} & 0.760\textsubscript{±0.017}\\
        \cline{1-4}
    \end{tabular}
    \label{tab:pmp}
    \vspace{-0.1cm}
    
\end{table}

\subsubsection{Importance of Prompt Model Performance}
Table \ref{tab:pmp} shows the results of the experiments on the prompt model performance module, after the prompt speech quality selected using the positive selection of ExpPro, we compare the performance of the Top \( k \) and Bottom \( k \) prompts candidates to validate the role of prompt model performance module. The experimental results indicate that the module significantly enhances the user's listening experience.

\section{Conclusions and Future Work}

In this paper, we proposed ExpPro, a novel two-stage emotion prompt selection strategy that evaluates both the emotional quality of prompts and their generation performance.

ExpPro also performs dynamic prompt selection based on the input text to select the most relevant prompt among the emotional prompt candidates.
The experiments show that, compared to the baseline methods, the speech generated using the prompt selection strategy proposed in this paper demonstrates advantages in emotional expressiveness, perceptual quality, and content accuracy. In the future, we will further explore prompt selection strategies across other dimensions and try to apply them to various tasks such as text-to-audio.

\section{Acknowledgements}
This work was supported by the National Natural Science Foundation of China under Grant U23B2053 and Grant 62176182.

\newpage

\printbibliography

\end{document}